# Increased antimicrobial activity of ZnO nanoparticle thin films: Effect of surface structuration

Imroi El-Habib[a,b,1], Sarah Dine[a], Alex Lemarchand[a], Christine Mielcarek[b], Mamadou Traore[a], Rabah Azouani[b,*]

[a] Laboratoire des Sciences des Procédés et des Matériaux (LSPM-CNRS UPR 3407), Institut Galilée, Université Sorbonne Paris Nord, 99 Avenue Jean-Baptiste Clément, 93430 Villetaneuse, France

[b] Ecole de Biologie Industrielle (EBI), 49 Avenue des Genottes – CS 90009, F95895 Cergy Cedex, France

ABSTRACT

This work explores the antimicrobial efficacy of ZnO nanoparticles (NPs) films with varying microstructure, crystallographic orientation and surface roughness. Thin films with different microstructures were obtained by varying precursor concentrations (0.1 M, 0.25 M, and 0.4 M) and solvent types (Ethanol, Propanol, and Butanol). The characterization of the films by X-ray diffraction (XRD) revealed films with varying degrees of orientation along the c-axis, which increased with the concentration of the precursor and with the decrease in the carbon chain length of the alcohol. Raman spectroscopy analysis confirmed the formation of a hexagonal wurtzite ZnO. Scanning Electron Microscopy (SEM) analysis revealed the formation of dense and homogeneous films with a thickness of 93 nm. Atomic force microscopy (AFM) showed changes in surface irregularity and an increase in surface roughness with the carbon length of solvent and a maximum roughness at a precursor concentration of 0.25 M in propanol. The antimicrobial activity was assessed following ISO 22196:2011 standards with few modifications. Results indicated that ZnO NPs films with rougher surfaces reduced *coli* (gram(-)), *Pseudomonas aeruginosa* (gram(-)) and *Staphylococcus aureus* (gram(+) ) bacteria populations due to an increase of contact surface between NPs with bacteria. Their effectiveness against *Candida albicans* was moderate, with a maximum reduction of 0.72 $\text{Log}_{10}$ $\text{CFU/cm}^2$ with rougher film.

**Introduction**

Microbial infectious diseases pose a significant global health concern, with an escalating economic [1,2] and social [2] burden. Most life-threatening microbial infections are acquired in healthcare settings, such as hospitals and intensive care units, commonly called Nosocomial Infections or Healthcare-Associated Infections (HCAIs) [2]. Microbial infections and their manifestations disrupt every stage of medical methodologies, from the formulation of misleading and erroneous diagnoses to resulting in detrimental surgeries and unsuccessful and incomplete treatments [3]. The severity of these conditions is exacerbated by the presence of additional or central infections [3]. Present challenges faced by healthcare systems are primarily attributed to multi-drug resistance (MDR), which constitutes a severe threat to public health [4]. Consequently, antimicrobials have become a crucial and fundamental component of preventive measures.

Nanomaterials are increasingly garnering attention for their potential use as antimicrobial agents. Their physicochemical and structural properties undergo modification [2], [5], [6], [7], [8], [9], primarily driven by surface features and quantum size effects [7], [8], [9], which enhance their activity compared to their bulk counterparts. In particular, their attractive antimicrobial properties are due to the increased specific area and enhanced particle surface reactivity [10], [11].

---

[1] Corresponding authors.
E-mail adresses : el-habib.imroi@lspm.cnrs.fr (I. El-Habib), r.azouani@hubebi.com (R. Azouani)

ZnO has been recognized as a versatile and promising inorganic material suitable for various applications. It is classified as inexpensive, and biocompatible with human cells [12], [13], [14]. More importantly, it exhibits antimicrobial activity, making it particularly relevant in this study [15], [16], [17], [18]. ZnO is classified as safe by the US Food and Drug Administration (FDA)( 21CFR182.8991) [19]. It has been utilized in various commercial dermatological applications such as cream, lotions, ointments and mouthwashes in order to prevent pathogens growth [2]. Additionally, it is used as food preservative [20].

However, the use of ZnO nanoparticles in suspension as antibacterial agents has been shown to have adverse effects on human health [2], [21]. Major concerns in human health include toxicity [22], [23], [24], biodiversity, and bioaccumulation[24], [25], [26]. In environmental contexts, nanoparticles accumulate in ecosystems, presenting a threat to living organisms[25], [27]. Thus thin ZnO nanoparticles on a substrate can be a promising candidate for antibacterial application, if it retains their antibacterial properties.

ZnO thin films can be prepared using numerous techniques, including sol-gel spin coating, sol-gel dip coating, aerosol-assisted chemical vapor deposition, magnetron sputtering, pulsed laser deposition (PLD), and atomic layer deposition (ALD) [28]. Among these techniques, sol-gel dip-coating has attracted much attention due to its low cost, minimal equipment requirements, fast processing, and ease of control of processing parameters. The sol–gel process has numerous advantages in the fabrication of ZnO films, including strong c-axis alignment, large area coverage and a lower annealing temperature [20].

Existing literature has shown the influence of parameters such as doping rate [29], [30], [31], precursor concentration [32], Zn/O ratio [33], morphology and size of ZnO thin films [34] on their antibacterial activity. Some authors suggest that triangular nanosheets have the highest biocidal activity, while others indicate that cubic NPs are the most effective due to their exposed planes [35]. Numerous studies have demonstrated that preferential solid growth of ZnO nanocrystals along the c-axis (002) plane improves their final properties [36], [37], [38]. A higher orientation of the c-axis (002) peak indicates minimization of the films' internal stresses and free surface energy, as well as dense atomic packing [38]. However, few reports focus on the influence of the surface structuring of ZnO thin films on their antimicrobial performance. The surface structure determines the contact area and adhesion strength between the bacterial cell and the surface [39]. Features such as hydrophobicity [40], potential and charge [41], patterning [42] and roughness [43], [44] have been shown to have an impact on microbial adhesion. Yoda et al. [45] found that S. aureus adhesion was greater on coarse than on fine surfaces. Silva et al. have shown that roughness is the main factor determining bacterial adhesion [46]. In general, surfaces with moderate wettability are better able to bind bacteria or cells than extremely hydrophobic or hydrophilic surfaces [47]. Li et al. [48] studied bacterial adhesion to metal oxide surfaces. They showed that bacterial adhesion is strongest on the most hydrophobic surfaces and increases with ionic strength. Such features on ZnO films, in addition to ZnO's intrinsic antimicrobial activity, could enhance their antimicrobial activity. Within this framework, the study of the influence of the microstructure of ZnO films on their antimicrobial activity will be presented.

Therefore, this study aims to explore the impact of the microstructural of ZnO thin film on their antimicrobial activity against Gram-negative, Gram-positive, and fungi according to ISO 22196:2011. ZnO thin films coated onto glass were prepared using the sol-gel dip-coating method. The influence of precursor concentrations and solvent type was investigated to prepare thin films with various microstructures.

**Material and Methods**

**Material**

Zinc acetate dihydrate (≥ 99%), copper acetate monohydrate (98%), butan-1-ol (≥ 99.5%), propan-1-ol (≥ 99.5%), ethanol (≥ 99.5%), acetone (≥ 99.5%), and silicone oil were from Sigma-Aldrich. Isopropanol (≥ 99.5%) was obtained from Acros Organics. Monoethanolamine (MEA) (100%) was acquired from Emprove. Eugon LT Sup is from. Microscope slide is from VWR. Microscope glass slide are from.

**Methods**

Elaboration of ZnO thin films

Solutions at concentrations of 0.1 M, 0.25 M and 0.4 M were prepared by dissolving the required amount of zinc acetate dihydrate [$Zn(CH_3COO)_2, 2H_2O$] in 100 ml propanol for all three concentrations, and in butanol and ethanol for the 0.25 M concentration. These solutions contained monoethanolamine (MEA) as a chelating and stabilizing agent, with a $Zn^{2+}$/MEA molar ratio of 1:1. After stirring the mixture at 60°C for 2 hrs for ZnO to allow hydrolysis and condensation reactions to take place and obtain transparent, homogeneous solutions, the solutions were left to age for 24 hrs in a glovebox ($O_2$ and $H_2O$ <0.5 ppm) before being used as a coating solution. The glovebox, where oxygen and water levels are low, was used to prevent precipitation of colloidal solutions. ZnO thin films were deposited on standard glass substrates (FEA, Microscope slides ground Edges Plain, FEA Industries Inc., Morton, PA, US) measuring 76 x 25 mm. Prior to film preparation, the glass substrates were immersed in a 98% sulfuric acid ($H_2SO4$) solution for 4 days, rinsed with distilled water three times until the pH reached 7, then dried in an oven to ensure good adhesion of the thin films. Sulfuric acid treatment of glass substrates hydrolyzes the -Si-O-Si-O-Si- groups in the substrate, increasing the surface density of the Si-OH functions required to form covalent bonds between ZnO and the substrate. The films were deposited using the dip coating technique at a deposition rate of 5 cm/min. The total number of layers deposited was 5. Between each deposition of the first three layers, the samples are left to dry in the glove box for 24 hours to avoid contamination. The purity and homogeneity of these first three layers are crucial to ensure good adhesion of the last two layers. For the 4th and 5th layers, the samples are oven-dried for 5 minutes at 200°C. Finally, the samples are annealed at 500°C to allow the formation of crystalline zinc oxide.

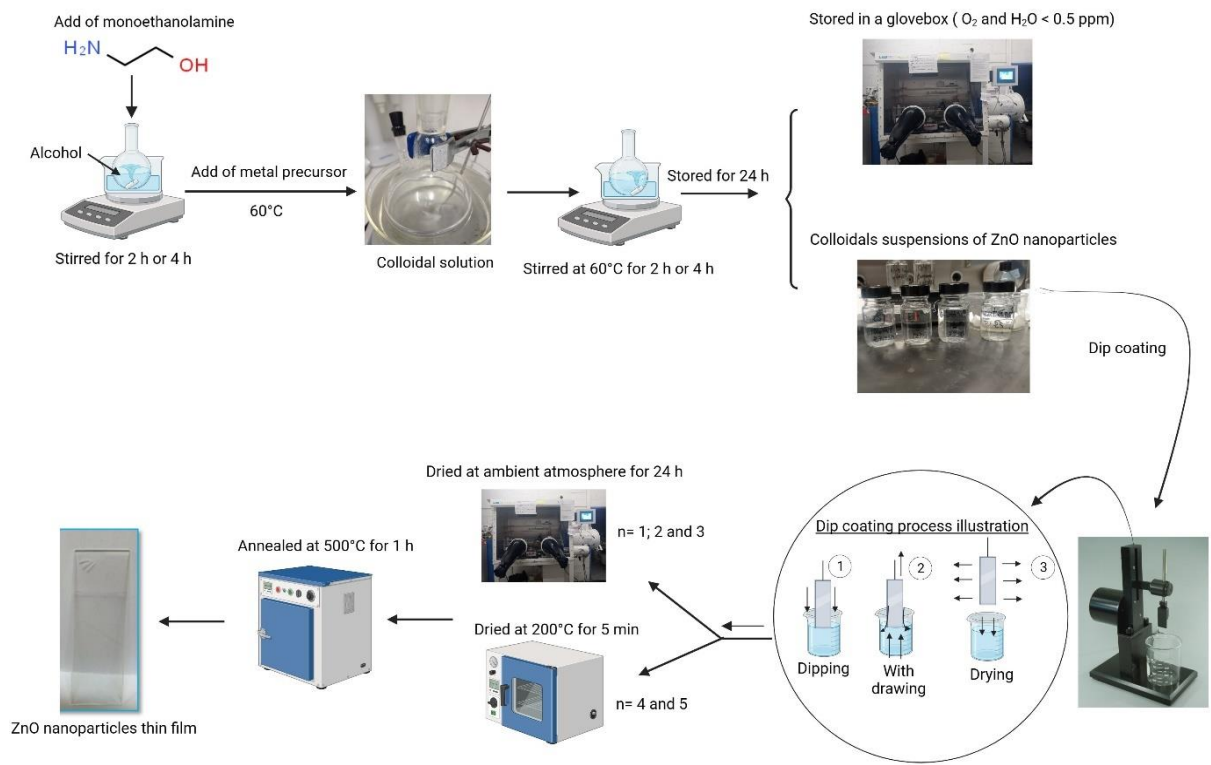

**Figure 1**: Schematic illustration of the thin film elaboration process, where 'n' represents the number of layers on the glass slide.

Samples characterization

The crystallinity and phases of the samples were determined using an X-ray diffractometer (INEL XRG 3000) equipped with a monochromatic cobalt source ($\lambda$ K$\alpha$1 (Co) = 1.788976 Å). Surface morphology was examined using scanning electron microscopy (SEM) (HITACHI TM 3000), while vibrational properties were assessed via Raman spectroscopy (HORIBA Jobin-Yvon HR800) with an excitation wavelength of $\lambda$ = 633 nm. AFM imaging was carried out in tapping mode using a Digital Instrument IIIA 3100 nanoscope microscope equipped with silicon nitride tips with a radius of curvature of around 15 nm.

Antimicrobial activity

The antimicrobial activity of ZnO thin-film was evaluated against gram-negative *Escherichia coli* (ATCC 8739) and *Pseudomonas aeruginosa* (ATCC 9027), gram-positive bacteria *Staphylococcus aureus* (ATCC 6538), and the pathogenic yeast *Candida albicans* (ATCC10234). Assessment of the antimicrobial activity of ZnO thin films was conducted following ISO 22196:2011 standards with minor modifications.

Strains were grown on nutrient agar (containing 5.0 g meat extract, 10.0 g peptone, 5.0 g sodium chloride and 15.0 g agar powder) at 35°C for *E. coli* and *S. aureus*, and at 30°C for *P. aeruginosa* and *C. albicans*. Bacteria were cultured for 24 hours and fungi for 72 hours before being subcultured. Subsequently, the strains were transferred to flasks containing 1/500 nutrient broth (3.0 g of meat extract, 10.0 g of peptone, and 5.0 g of sodium chloride), and their initial concentrations were adjusted to 8 Log UFC/mL for bacteria and 7 Log UFC/mL for fungi by measuring their optical density at 660 nm using a laboratory-established calibration method.

The final concentration of microorganisms at 6 Log CFU/mL was chosen for the experiment. Microbial suspensions (0.1 mL) were spread onto thin films, and the samples were covered with a polypropylene cover film (2 x 5 cm$^2$) to ensure contact between thin films and the samples. The samples were then incubated at 35°C with a relative humidity of ≥ 90% for 24 hours. Following incubation, microbial cells were recovered from ZnO thin films by placing the samples into 10 mL of Eugon LT broth and 10 g of 1 mm glass beads, then vortexing for 1 minute.

The antimicrobial activity was evaluated by colony counting on agar plates. Positive controls were conducted using glass substrates treated with sulfuric acid under the same conditions as the samples. The concentration of the bacterial population deposited on the positive control immediately after inoculation was measured to determine the actual concentration deposited. All tests were performed in triplicate, and the antimicrobial activity was calculated using the following formula:

$R = U_t - A_t$  in Log CFU/cm$^2$ (1)

Where,

R : is the antibacterial activity

$U_t$ : is the average of the common logarithm of the number of viable bacteria, in Log CFU/cm$^2$, recovered from the positive control after 24 h;

$A_t$ : is the average of the common logarithm of the number of viable bacteria, in Log CFU/cm$^2$, recovered from NPs thin film after 24 h.

The number of viable microorganisms was determined according to the following equation:

$N = (100 \times C \times D \times V)/A$  in CFU/cm$^2$ (2)

Where,

N : is the number of viable microorganisms recovered per cm$^2$ per test sample

C : is the average plate count for the triplicate plates

D : is the dilution factor for the plates counted;

V : is the volume, in ml, of Eugon LT Sup added to the sample;

A : is the surface area, in mm$^2$, of the cover film.

**Results and discussions**

**Structural properties**

 **Figure 2** shows X-ray diffractograms of samples prepared from different precursor concentrations in propanol (**Figure 2**-**a**) and various solvents (**Figure 2**-**b**). All samples are well crystallized in the Wurtzite structure of ZnO, with no secondary phase appearing.

Samples grown at different concentrations in propanol show a strong orientation primarily along the c-axis ( **Figure 2**-**a**). ZnO crystallites were generally thought to cluster along the c-axis (002) due to their lower surface free energy [49].  The intensity of the peaks increases with the concentration, consistent with findings in the literature [50]. This is attributed to faster growth, which reduces the likelihood of the material rearranging into a more defined structure. The degree of orientation of the films was assessed using the texture coefficient (TC), calculated from the following formula and displayed in **Figure 3** :

CT values are less than 1, except in the (002) plane, confirming the preferential orientation of films at different concentrations. The film from the highest concentration has the highest c-axis TC value, signifying a higher degree of orientation along the c-axis and better crystallinity.

XRD diffractograms of films from different solvent types at the same zinc precursor concentration reveal the formation of hexagonal Wurtzite-crystallized zinc oxide (**Figure 2**-**b**).

The solvent type plays a crucial role in the crystalline orientation, as samples from propanol and ethanol show a preferential orientation along the c-axis, while that from butanol does not. Texture coefficient analysis further underscores the influence of solvent type, as it highlights the orientation of the (002) plane on the three samples from three different solvents, with their value being more significant than 1. The evolution of the degree of preferential orientation along the c-axis as carbon chain length decreases is also a direct result of the solvent type. These results are consistent with those reported in the literature [51], [52]. The high viscosity of long-chain solvents, such as butanol, slows down solvent evaporation and particle growth, resulting in a less ordered arrangement of the material on the films.

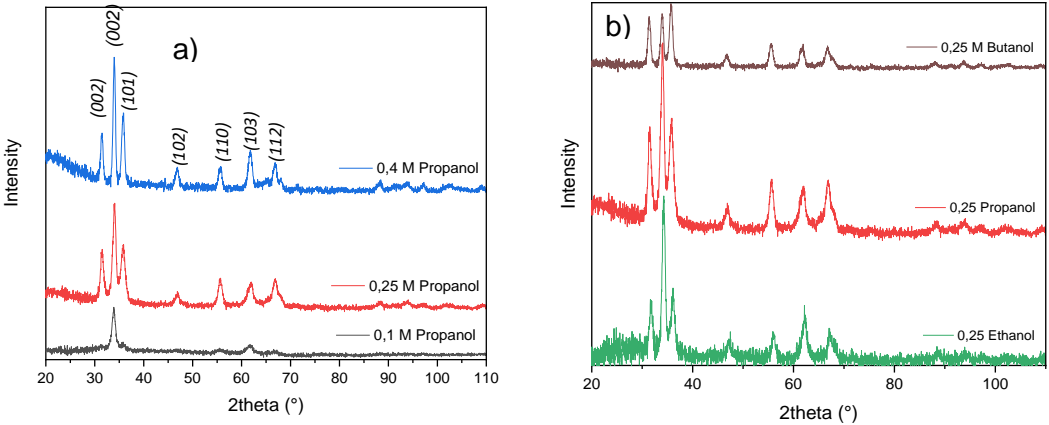

**Figure 2** : XRD diffractograms of ZnO thin films: a) Influence of concentration and b) Influence of solvent type.

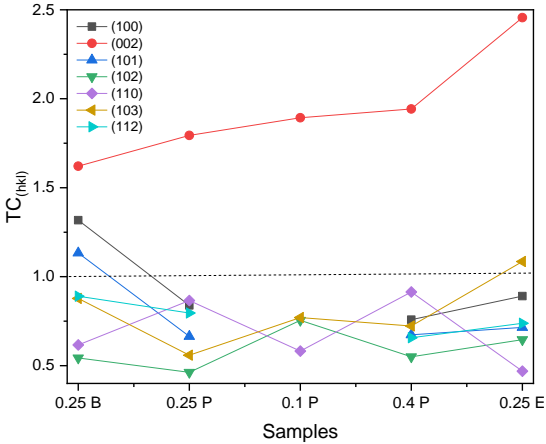

**Figure 3** : Texture coefficient as a function of precursor concentration and solvent type: 0.25 B for 0.25 M Butanol; 0.25 P for 0.25 M Propanol; 0.1 P for 0.1 M Propanol; 0.4 P for 0.4 M Propanol and 0.25 E for 0.25 M Ethanol.

**Raman analysis**

**Figure *4*** shows the Raman spectra of ZnO thin-film samples processed in propanol. According to the irreducible representation, ZnO has eight phonon modes: Γopt = A1 + E1 + 2E2 + 2B1 [16]. The A1 and E1 modes are polar, active in both the Raman and infrared, and are divided into transverse optical phonons (TO) and longitudinal optical phonons (LO). The two B1 modes are neither active in the Raman nor the IR (silent). The two E2 modes (E2 low and E2 high) are non-polar and active only in the Raman. The peaks observed at 435 cm$^{-1}$ are due to the Raman-active E2 optical phonon mode of ZnO, confirming a hexagonal wurtzite crystal structure in line with the DRX, as mentioned in the earlier analysis. Naturally, the intensity of these peaks increases with concentration due to more nanoparticles. The firm peaks at 565 cm$^{-1}$ were attributed to the A1(LO) mode [53]. These peaks could

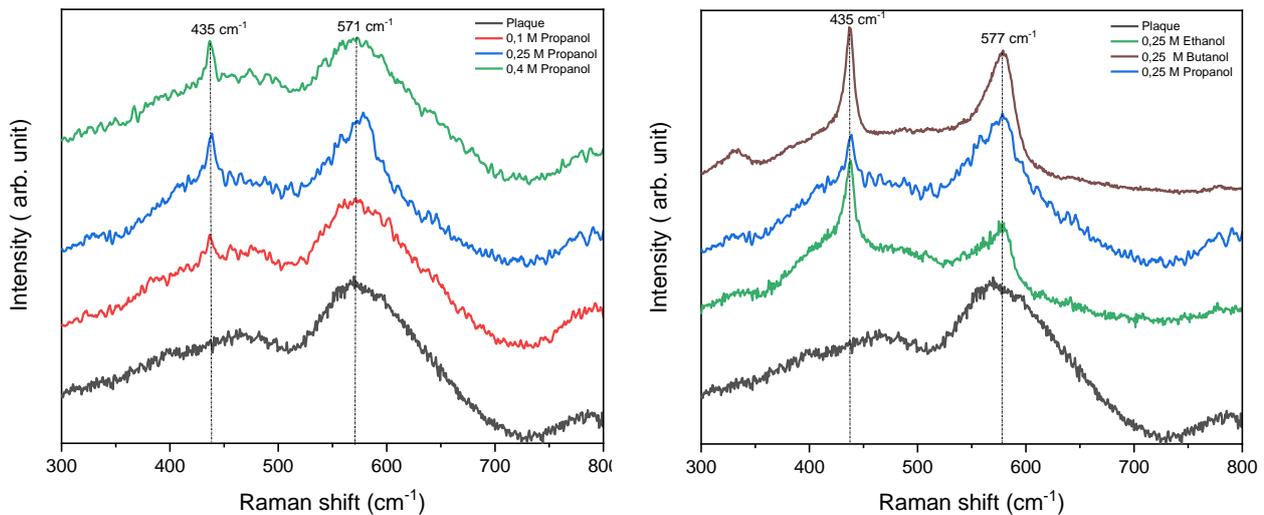

be due to the formation of defects such as zinc interstitials and oxygen vacancies [53].

**Figure 4** : Raman spectra of ZnO thin films.

**Morphological studies**

**Figure 5** shows the surface area and thickness of the nano deposit produced from a 0.4 M solution of ZnO in propanol. The films are continuous and homogeneous, consisting of ZnO nanocrystallites with a few agglomerates. The film thickness is 93 nm.

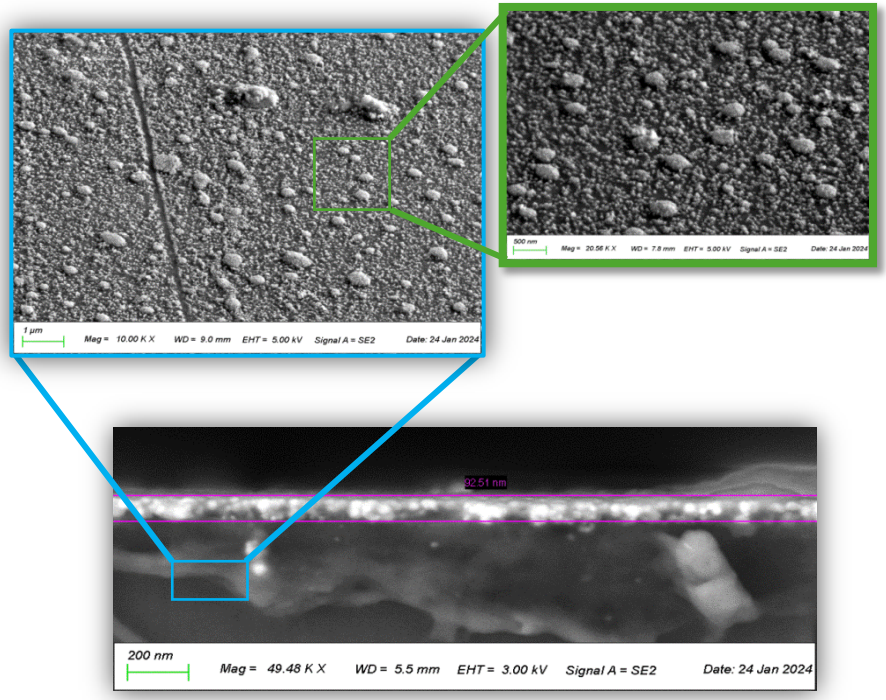

**Figure 5** : Scanning electron microscopy of ZnO films made from a 0.4 M precursor solution in propanol.

**Figure 6** shows ZnO thin films deposited on glass substrates.

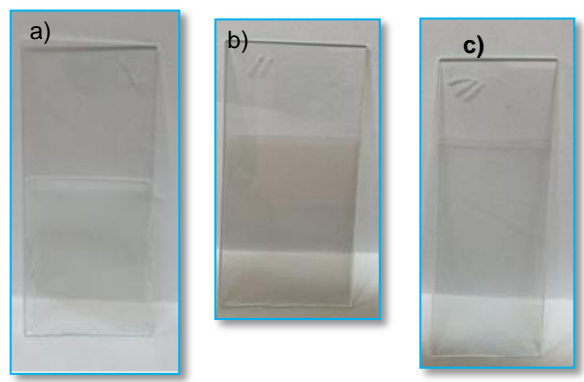

**Figure 6** : ZnO thin films: a) 0.1 M Propanol; b) 0.25 M Propanol and c) 0.4 M Propanol.

**Surface topography**

Two-dimensional AFM images of the samples are shown in **Figure 7**. The influence of the concentration and type of synthesis solvent on surface roughness is shown in Table 1. The surface of the sample deposited in the 0.4 M solution in propanol was the smoothest, with a roughness (Ra) of around 6.55 nm. The surface roughness of the sample deposited at 0.1 M in propanol was almost similar to that of the sample deposited at 0.4 M in propanol, with a roughness of 8.28 nm. However, when the concentration of the precursor in propanol was 0.25 M, the roughness of the samples was significantly altered. The grains are much larger, and grain boundaries are more numerous. The surface of the samples is rougher than that of samples deposited at 0.1 M and 0.4 M, exhibiting a wide variation in height, as shown in **Table 1**. Therefore, the concentration of the zinc precursor can significantly alter the structure and morphology of the surface, thereby increasing grain size.

The surface structure and morphology change was evident when the synthesis solvent was changed. The surface of the ZnO film became much rougher, and the variation in height became much more significant in the sample from butanol. The ethanol sample has the smoothest surface.

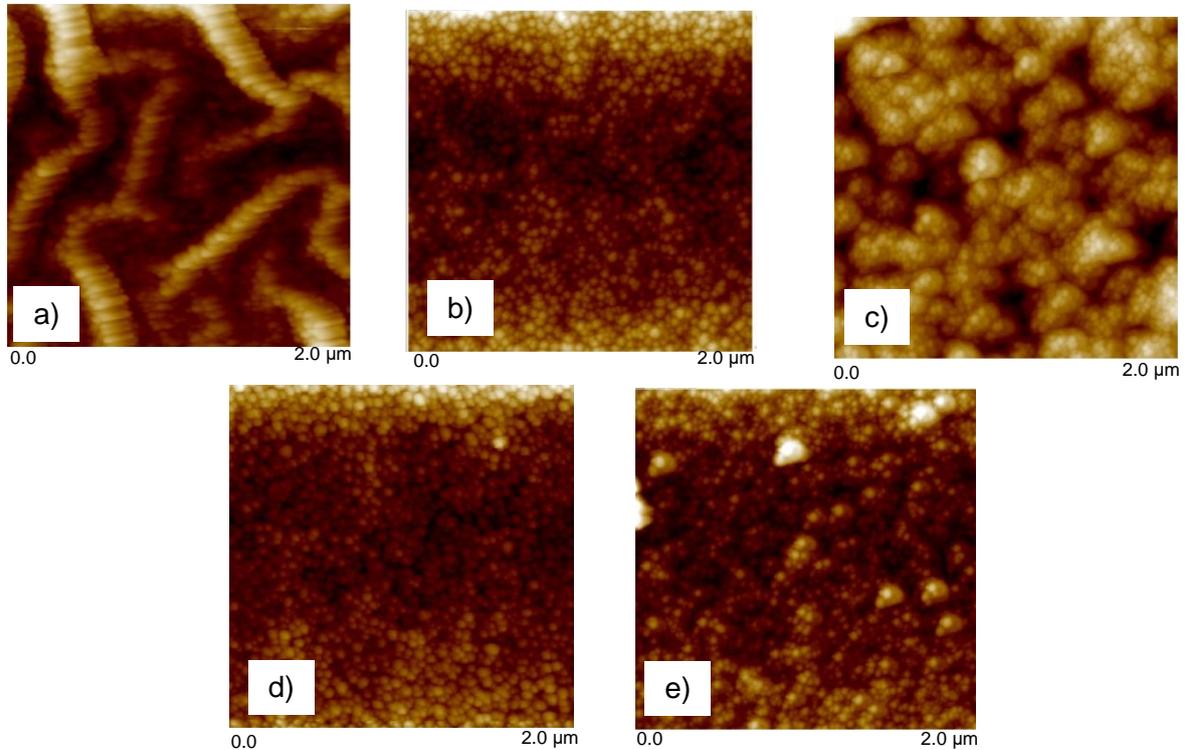

**Figure 7** : Two-dimensional AFM image of samples: a) 0.25 M Butanol, b) 0.1 M Propanol, c) 0.25 M Propanol, c) 0.4 M Propanol, and e) 0.25 M Ethanol.

**Table 1** : Effect of zinc precursor concentration and solvent type on the roughness of ZnO films.

| Samples | Ra (nm) | Rq (nm) | Rmax (nm) |
|---|---|---|---|
| 0.25 M Butanol | 14.3 | 17.4 | 96.8 |
| 0.1 M Propanol | 8.28 | 10.3 | 65.6 |
| 0.25 M Propanol | 12.3 | 15.6 | 12.9 |
| 0.4 M Propanol | 6.55 | 8.32 | 55.2 |
| 0.25 M Ethanol | 6.91 | 8.70 | 62.9 |

Three-dimensional images showed that ZnO thin films were grains growing vertically, forming columnar structures along the c-axis, perpendicular to the substrate surface (**Figure 8**). The columnar structure of the ZnO film is generally oriented along the c-axis, and its upper surface tends to be polar with high surface energy [54]. Irregularities in surface topography were observed as the length of the alcohol carbon chain increased.

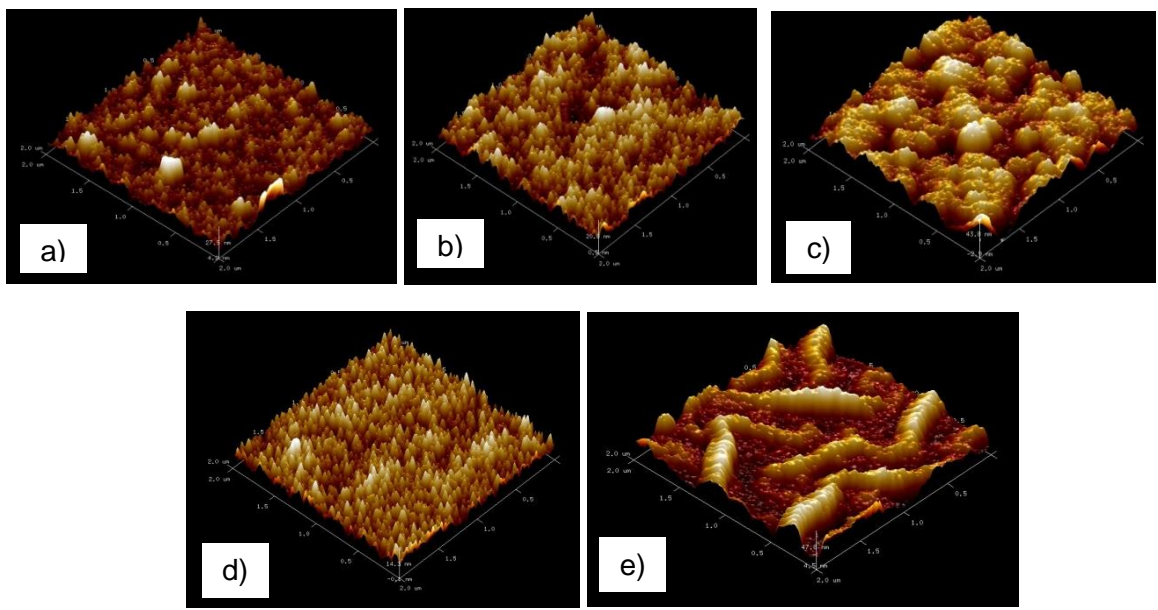

**Figure 8** : Three-dimensional AFM image of samples a) 0.25 M Butanol; b) 0.1 M Propanol; c) 0.25 M Propanol; c) 0.4 M Propanol and e) 0.25 M Ethanol.

**Antimicrobial activity**

**Figure 9** illustrates the antimicrobial effect of ZnO films. The absence of colonies on the Petri dishes indicates an inhibitory effect on the samples. The results show that the 0.1 M Propanol, 0.25 M Propanol, 0.4 M Propanol, 0.25 M Ethanol and 0.25 M Butanol samples have an inhibitory effect on *E. coli*. The same observation applies to other bacteria, except the 0.4 M Propanol sample, which inhibits *S. aureus* and *P. aeruginosa* to a lesser extent.

Evaluation of the 0.1 M Propanol, 0.25 M Propanol and 0.4 M Propanol samples on *C. albicans* reveals that the 0.25 M Propanol sample has a greater inhibitory effect than the other two samples (**Figure 10**).

The microstructure-dependent improvement in film inhibition can be explained by the difference in film morphology and, more likely, film roughness.

Moreover, the interaction between bacteria and patterned surfaces depends on the geometry, size and height of the patterned surface [39]. Bohinc et al. [55] studied different levels of roughness (0.07 µm, 0.58 µm, 0.99 µm, 2.5 µm and 5.8 µm) and observed that the adhesion rate of *E.coli*, *P.aeruginosa* and S.aureus increased in proportion to surface roughness. High roughness increases the contact surface between bacteria and ZnO films, thereby increasing the likelihood of NPs interacting with and killing cells. The impact of roughness can be explained by the fact that when roughness is relatively high (Ra = 6-30 nm), bacteria prefer to bind to the rougher surface [39]. On the other hand, when roughness is between 0.23-6.13 nm, the opposite effect is observed, i.e. bacteria adhere to the smoothest surface under static culture conditions [39]. Thus, the 0.4 M Propanol sample, having a relatively low roughness (Ra = 6.91 nm), would allow less adhesion of S. aureus and *P. aeruginosa* bacteria, thus reducing its inhibitory effect. Moreover, the roughest sample in the series tested on *C. albicans* had the highest antifungal activity.

The different samples had the same antimicrobial effect on *E. coli*, probably due to its rod shape and larger diameter, optimizing the contact surface between the bacteria and the films.

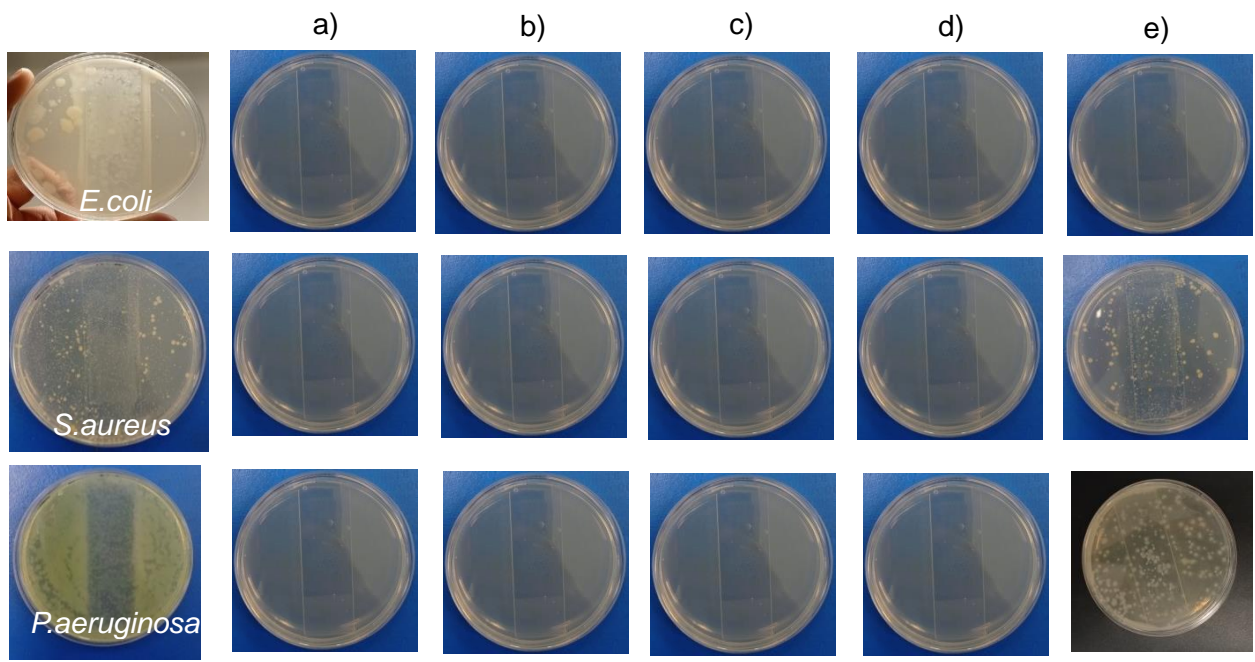

**Figure 9** : Antibacterial activity of ZnO thin films: a) 0.1 M Propanol; b) 0.25 M Propanol; c) 0.25 M Ethanol; d) 0.25 M Butanol and e) 0.4 M Propanol.

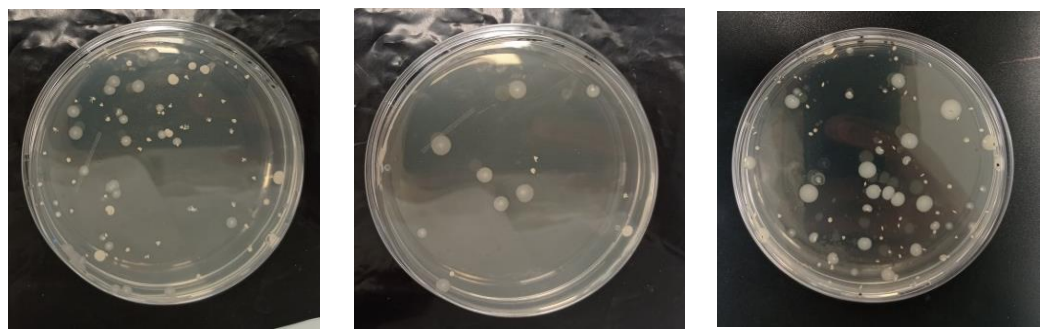

**Figure 10**: Antifungal activity of ZnO thin films against C.albicans: a) 0.1 M Propanol; b) 0.25 M Propanol and c) 0.4 M Propanol.

| Bacteria | Control (Ut) | 0.1 M Propanol | 0.25 M Propanol | 0.25 M Ethanol | 0.25 M Butanol | 0.4 M Propanol |
|---|---|---|---|---|---|---|
| *E.coli* | 6.63 ± 0.10 | 6.63 | | | | |
| *S.aureus* | 5.84 ± 0.16 | 5.84 | | | | 4.09 |
| *P.aeruginosa* | 625 ± 0.24 | 6.25 | | | | 4.61 |

**Table 2** presents the films' antibacterial activity, which confirms the results presented previously. The 0.4 M propanol sample showed similar activity on *S. aureus* and *P. aeruginosa*.

**Table 2** : Antibacterial activity of ZnO (R) films in $\text{Log}_{10}$ UFC/cm$^2$.

Table 3 shows the antifungal activity of ZnO films on *C.albicans*. Antifungal activity is highest for the 0.25 M propanol sample, with the highest roughness, confirming the above results. According to regulations on the antimicrobial effect of surfaces, a surface has an antimicrobial effect from a logarithmic reduction of at least 0.5 $\text{Log}_{10}$ UFC/cm².

**Table 3** : Antifungal activity of ZnO (R) thin films in $\text{Log}_{10}$ UFC/cm$^2$.

| Microorganism | 0.1 M Propanol | 0.25 M Propanol | 0.4 M Propanol |
|---|---|---|---|
| *C.albicans* | 0.41 | 0.72 | 0.22 |

**Conclusion**

In this study, we investigated the correlation between the structural properties of ZnO films and their antimicrobial activity. Antimicrobial activity increased with film roughness, due to an increase in the contact surface between NPs and bacteria. Total inhibition (R=$U_t$) of bacteria was observed for the roughest samples. A significant and promising reduction (R= 0.72) in candida cells was observed on the ZnO sample with the highest roughness, underlining the potential impact of this research on healthcare.

Thus, modified nanotopographies create a physically hostile surface for bacteria, killing cells through biomechanical damage [35]. These results pave the way for the design of ultra-biocidal and durable nanoparticle-based surfaces.